%% file: tj-nc.tex
\documentclass[a4paper,10pt,leqno,twoside]{tobart}

\usepackage[english]{babel} \usepackage{inputenc, amsmath, amssymb, latexsym,
  epic, epsfig, rotating, fancyheadings, amsthm, pifont, empheq}

\input{bibstandard}

\newtheoremstyle{tobthm}{3pt}{3pt}{\itshape}{0pt}{\bfseries}{.}{0.5eM}{}
\theoremstyle{tobthm}

\newtheorem{definition}{Definition}[section]
\newtheorem{thm}[definition]{Theorem}

\newtheorem{prop}[definition]{Proposition}

\newtheorem{claim}[definition]{Claim}

\newtheoremstyle{tobrem}{3pt}{3pt}{\normalfont}{0pt}{\bfseries}{.}{0.5em}{}
\theoremstyle{tobrem}

\newtheorem{rem}[definition]{Remark}

\numberwithin{equation}{section}
\numberwithin{figure}{section}

\newcommand{\wh}{\ensuremath{\widehat}}

\title{\Large\textsc{Neuronal coding of pacemaker neurons -- A random dynamical
    systems approach}} \author{T.~J\"ager\thanks{Coll\`ege de France,
    Paris. Email: {\tt tobias.jager@college-de-france.fr}}}

\begin{document}

\setlength{\abovedisplayskip}{1.0ex}
\setlength{\abovedisplayshortskip}{0.8ex}

\setlength{\belowdisplayskip}{1.0ex}
\setlength{\belowdisplayshortskip}{0.8ex}

\maketitle 

\abstract{The behaviour of neurons under the influence of periodic external
  input has been modelled very successfully by circle maps. The aim of this note
  is to extend certain aspects of this analysis to a much more general class of
  forcing processes. We apply results on the fibred rotation number of randomly
  forced circle maps to show the uniqueness of the asymptotic firing frequency
  of ergodically forced pacemaker neurons. The details of the analysis are
  carried out for the forced leaky integrate-and-fire model, but the results
  should also remain valid for a large class of further models. }

\noindent
\section{Introduction}

Already in 1907, long before the molecular mechanisms of neural signal
transduction had been clarified, Louis Lapicque proposed a simple model for the
firing behaviour of a neuron
\cite{lapicque:1907,brunell/rossum:2007,abbott:1999}. A crucial feature of this
so-called integrate-and-fire model (IFM) is the separation of time-scales: the
stereotypical and extremely fast generation of an action potential is thought of
as being concentrated in a single moment of time, whereas the much slower
evolution of the membrane potential in the interspike intervals is modelled as a
continuous process. For many questions concerning the behaviour of neural
systems this level of abstraction turned out to be exactly the adequate one,
such that even nowadays, more than a hundred years after Lapicque's original
paper, the different variations of the IFM still play a central role in
theoretical neuroscience \cite{gerstner/kistler:2002}. One of their great
achievements was the explanation of so-called `paradoxical segments' that were
discovered in the experimental investigation of pacemaker neurons in the nervous
system of crayfish ({\em Procambarus clarkii}), sea slugs ({\em Aplysia
  californica}) and horseshoe crabs ({\em Limulus polyphemus})
\cite{perkeletal:1964,knight:1972b}. The counter-intuitive observation that was
made in these experiments was that an increase in the frequency of periodic
inhibitory presynaptic input can lead to an increase of the post-synaptic firing
frequency. This paradoxon was explained by relating the respective theoretical
models to monotone circle maps whose rotation number equals the ratio between
the input and the output frequency.  When the circle map has a stable periodic
orbit, then input and output frequency remain directly proportional on a small
neighbourhood (the paradoxical segment), disrespective of whether the input is
inhibitory or excitatory
\cite{perkeletal:1964,stein:1965,knight:1972a,segundo:1979} (see also
Section~\ref{Revisited}). Similar ideas have also been pioneered before by
V. Arnold in the study of cardiac cells \cite{glass:1991,arnold:1991}.

\begin{figure}[h] 
\begin{minipage}[h]{1.0\linewidth}
  \begin{center}
\epsfig{file=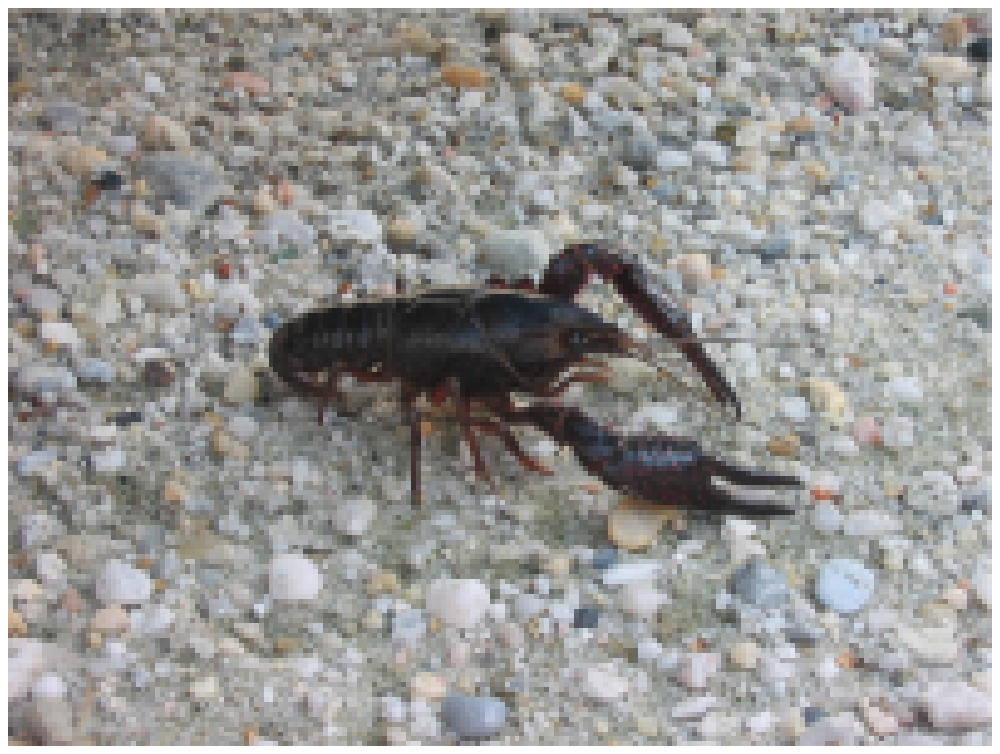, height=0.25\linewidth}
\epsfig{file=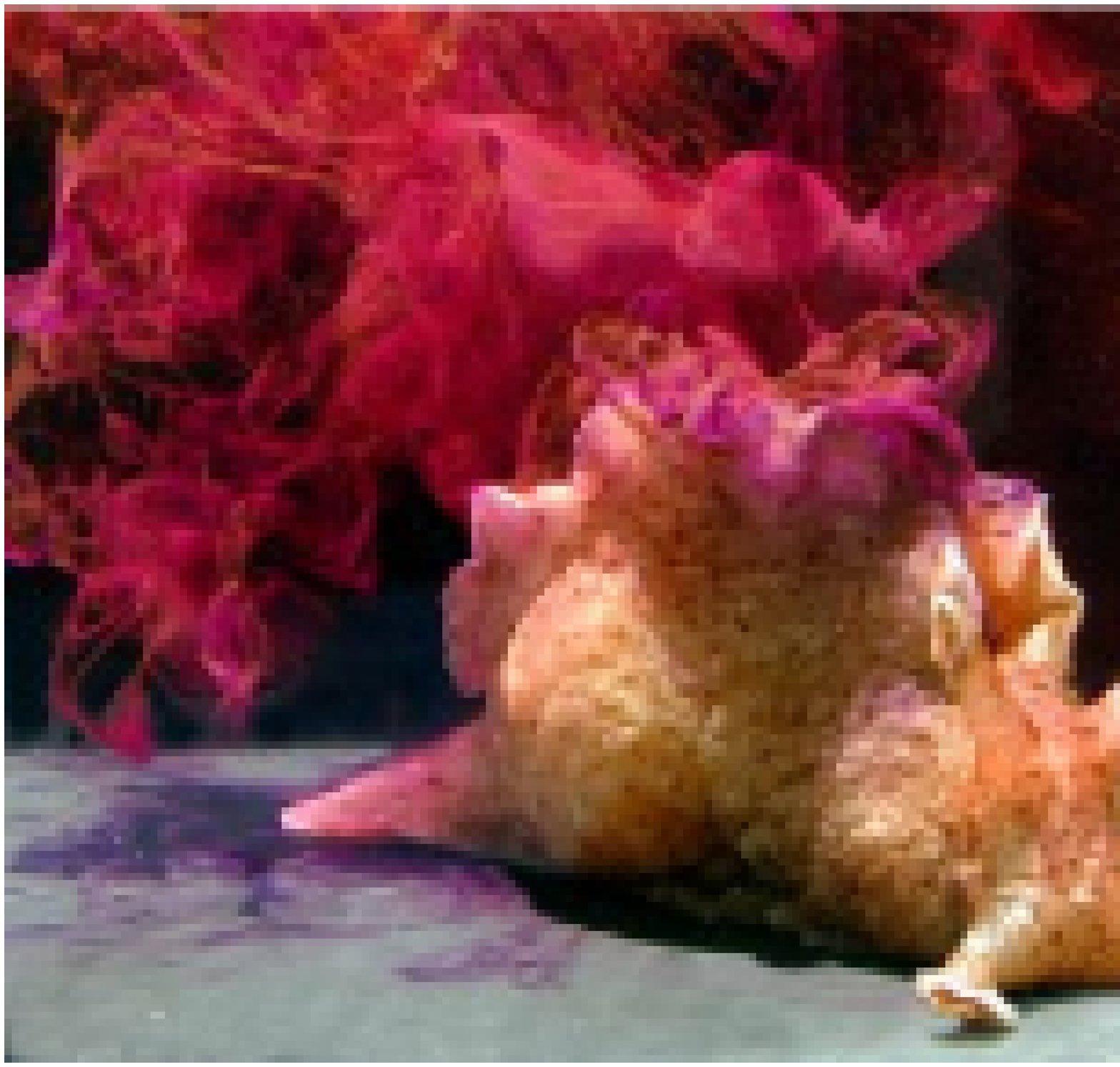, height=0.25\linewidth}
\epsfig{file=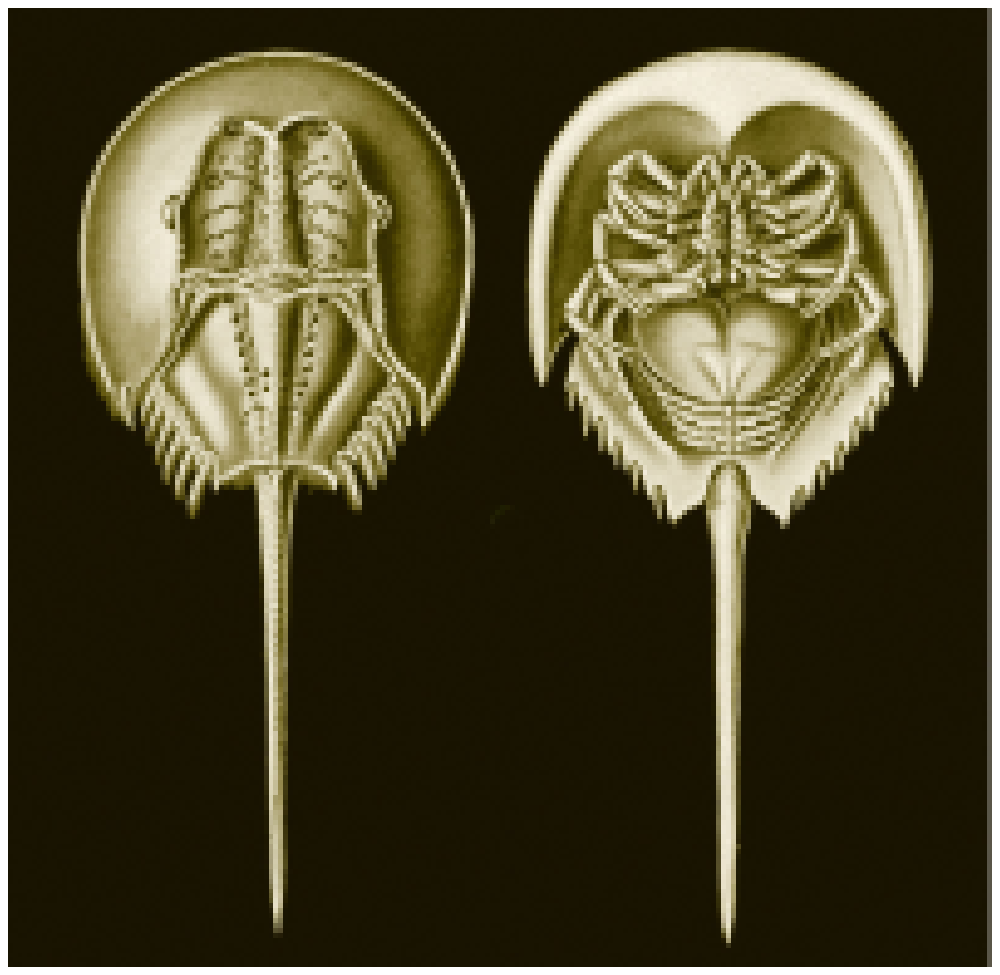, height=0.25\linewidth}
\end{center} {\caption{\label{fig1}\small Three species for which mode-locking
    phenomena in the nervous system have been investigated experimentally
    \cite{perkeletal:1964,knight:1972b}. From left to right: {\em Procambarus
      clarkii, Aplysia californica} and {\em Limulus polyphemus}
    \cite{pictures}.}}
\end{minipage}
\end{figure}

It must be said, however, that despite the great success of IFMs and the long
history of their investigation their rigorous mathematical description is still
restricted to a few very special situations. In particular, the only types of
external forcing that can be treated analytically so far are either periodic
\cite{pakdaman:2001,brette:2004} or stationary stochastic input
\cite{burkitt:2006a,lansky/ditlevsen:2008,vilela/lindner:2009}. Even the
superposition of the two -- noisy periodic input -- is mostly accessible only by
numerical methods \cite{burkitt:2006b}. Our goal here is take up the ideas used
in the analysis of the periodically forced IFM an to extend these to more
general forcing processes. Thereby, we restrict ourselves to deterministic
and/or random forcing, although it should be possible to adapt the approach to
models generated by stochastic differential equations as well. In order to state
the main results, we first recall the construction of the IFM.

The {\em membrane potential} $V(t)$ of a neuron ${\cal N}_1$ remains between a
lower threshold $V_l$ and an upper threshold $V_u$. $V(t)$ can never drop below
$V_l$ due to physiological constraints, whereas when it reaches $V_u$ the neuron
`fires', meaning that an action potential is triggered and $V(t)$ drops back to
a {\em rest potential} $V_r \in [V_l,V_u)$. Between the two thresholds, the
potential evolves according to an infinitesimal law
\begin{equation} \label{e.potential-law} \dot V(t) \ = \ F(t,V(t)) \ 
\end{equation}
with right side $F:\R^2 \to \R$ that should satisfy $F(t,V_l) \geq 0 \ \forall
t\in\R$. The dependence of $F$ on $t$ corresponds to the influence of external
time-dependent factors. The reset procedure when $V(t)$ reaches $V_u$ is usually
expressed as
\begin{equation} \label{e.setback} V(t^+) \ = \ V_r \quad \textrm{ if } V(t) =
  V_u \ ,
\end{equation}
where $t^+$ denotes the right-hand limit.  Identifying the interval $[V_l,V_u)$
with the circle $\kreis = \R / \Z$, this gives rise to a non-autonomous circle
flow (see Figure~\ref{fig1}). 

For fixed initial values $V(t_0)=x_0$, we denote by $t_n$ the time of the $n$-th
firing of the neuron ${\cal N}_1$. Then a very basic and fundamental question is
that of the existence and uniqueness of the {\em asymptotic firing frequency}:
under what assumptions does the limit
\begin{equation}
  \label{e.firing-frequency}
  \nu_{{\cal N}_1} \ = \ \nLim n/t_n \ 
\end{equation}
exist and when is it independent of the initial values $t_0$ and $x_0$?
\begin{figure}[h] 
\begin{minipage}[h]{1.0\linewidth}
  \begin{center} \epsfig{file=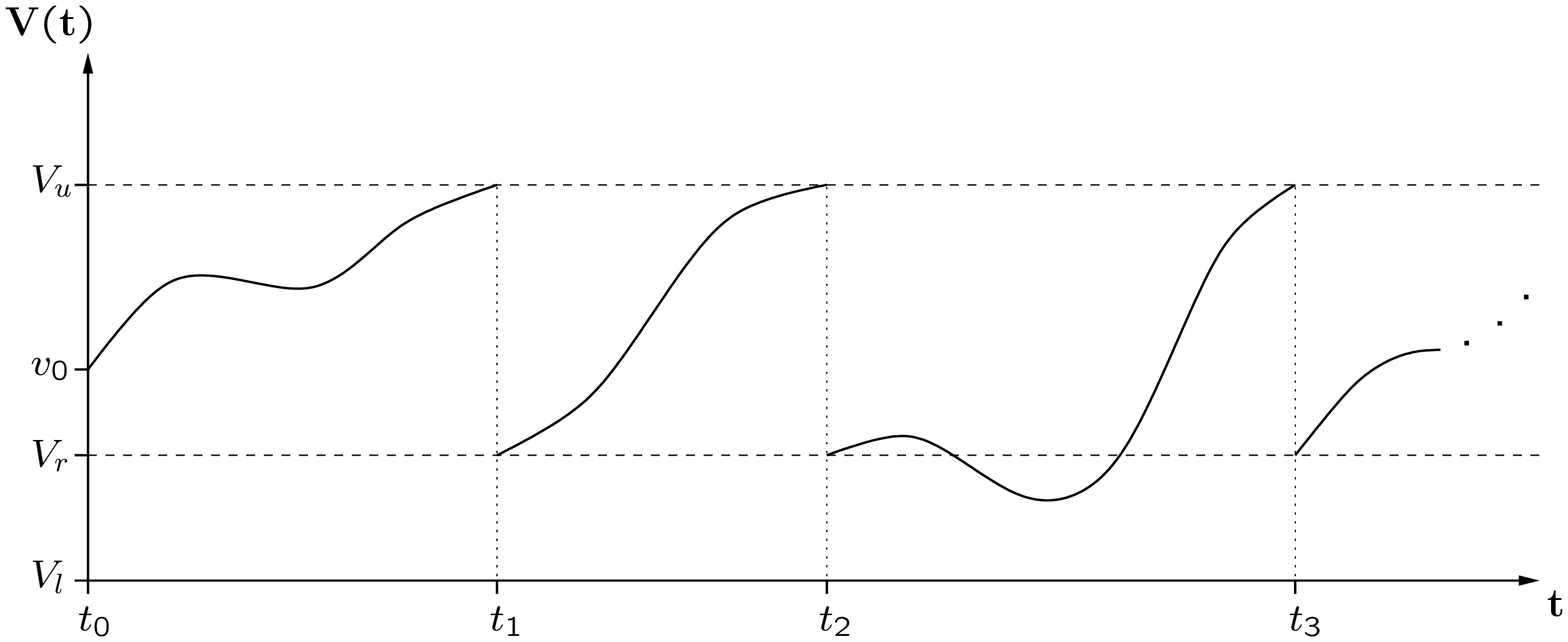, width=0.3\linewidth,
      angle=90}
  \end{center} {\caption{\label{fig2}\small Construction of the potential flow $V(t)$ generated by
    (\ref{e.potential-law}) and (\ref{e.setback}) with initial values $V(t_0)=x_0$.}}
\end{minipage}
\end{figure}

In the simplest case the external input is periodic in time with period
$p\in\R^+$. As mentioned, this situation is quite well-understood and has been
studied for a number of different versions of the IFM
\cite{pakdaman:2001,brette:2004}. The analysis depends on the choice of a
suitable Poincar\'e section for the flow, by which one obtains a circle map
$g:\kreis\to\kreis$ whose lift $G :\R \to \R$ generates the rescaled sequence
$t_n$, that is $t_{n+1}/p=G^n(t_n/p)$. (We briefly recall the construction in
Section~\ref{Revisited}.) Under suitable assumptions on the function $F$ this map
$g$ has good monotonicity properties that ensure the existence and uniqueness of
the rotation number
\begin{equation}
 \rho(G) \ = \ \nLim G^n(t_0/p)/n \ = \ \nu_{{\cal N}_1}^{-1}/p
\end{equation}
and thus of the asymptotic firing frequency. As indicated above, the existence
of a stable periodic orbit for the map $g$ yields an explanation for the
`paradoxical segments' in situations with inhibitory presynaptic input.
 
A good framework to study more general forcing processes is provided by the
theory of random dynamical systems, as exposed in \cite{arnold:1998}.  In order
to model the external input we will assume that there is an underlying {\em
  forcing process}, given by a {\em measure-preserving} or {\em metric dynamical
  system (metric DS)}, that is, a quadruple $(\Theta,{\cal B},\mu,\omega)$ where
$(\Theta,{\cal B},\mu)$ is a probability space and $\omega : \R \times \Theta
\to \Theta, (t,\theta) \mapsto \omega_t(\theta)$ and $\omega$ is a flow on
$\Theta$ that leaves the probability measure $\mu$ invariant (compare
\cite{arnold:1998}).  We usually write $\omega_t(\theta) = \theta\cdot t$. In
this setting (\ref{e.potential-law}) is replaced by
\begin{equation} 
  \label{e.Ftilde} \dot V(t) \ = \ F(\theta_0\cdot t,V(t))
\end{equation}
where $F : \Theta \times \R \to \R$ and $\theta_0$ is some initial value in
$\Theta$. In order to ensure that (\ref{e.Ftilde}) generates a RDS, we have to
impose some standard technical conditions on $F$. We say $F$ is {\em uniformly
  Lipschitz-continuous in $V$} if there exists a constant $L>0$ such that
\begin{equation}
\label{e.Lipschitz-condition} 
|F(\theta,V)-F(\theta,V')| \ \leq \  L\cdot |V-V'| \quad \forall \theta \in 
\Theta,\ V,V'\in\R \ .
\end{equation} 
For the moment, we will just assume that $F$ is bounded and uniformly
Lipschitz-continuous in $V$. These assumptions seem quite reasonable from the
physiological point of view. It is also possible to weaken them further to some
extent and we will discuss this issue in detail at the beginning of
Section~\ref{Construction}.

Due to the lack of a suitable structure on $\Theta$, it is not feasible anymore
to take a Poincar\'e section in this situation. However, it turns out that
instead the flow generated by (\ref{e.setback}) and (\ref{e.Ftilde}) can be
analysed directly by applying a result on the fibred rotation number of randomly
forced circle flows. (Basically, what we need is a slight modification of
statements from \cite{li/lu:2006} that we present in
Section~\ref{RDS-Rotnum}.) This leads to the following result.
\begin{thm}
  \label{t.asymptotic-ff} Suppose that the metric DS $(\Theta,{\cal
    B},\mu,\omega)$ is ergodic and the function $F$ in (\ref{e.Ftilde}) is
  bounded on $\Theta\times[0,1]$ and uniformly Lipschitz-continuous and
  non-increasing in $V$. Then the model described by (\ref{e.setback}) and
  (\ref{e.Ftilde}) has a unique asymptotic firing frequency in the following
  sense: There exists a real number $\nu$ such that for $\mu$-almost every
  $\theta_0\in\Theta$ and all $x_0\in [V_l,V_r)$ there holds
  \begin{equation}
    \nLim n/t_n \ = \ \nu \ .
  \end{equation}
\end{thm}
A well-known example to which this statement applies is the {\em leaky
  integrate-and-fire model} (LIFM), in which the function $F$ takes the form
\begin{equation}
  \label{e.lifm}
  F(\theta,V) \ = \ -a(\theta)\cdot V + b(\theta) \ . 
\end{equation}
Here $a :\Theta \to \R^+$ corresponds to
the membrane permeability%
\foot{Usually the membrane permeability is chosen to be fixed, but as we will
  see it presents no additional cost to assume that it is time-dependent as
  well}
whereas $b:\Theta \to \R$ reflects the external input. In order to apply
Theorem~\ref{t.asymptotic-ff} we have to assume that $a$ and $b$ are bounded.
Slightly weaker conditions are again discussed at the beginning of
Section~\ref{RDS-Rotnum}. The LIFM was first introduced by Stein
\cite{stein:1965} and then further investigated both theoretically and
experimentally by Knight \cite{knight:1972a,knight:1972b}. In contrast to the
so-called {\em `perfect integrator'} or {\em `perfect IFM'}
($\frac{\partial}{\partial V} F \equiv 0$) with `infinite memory' used by
Lapicque, it takes into account the exponential decay of the membrane
depolarisation in time after excitation.

\begin{figure}[h] 
\begin{minipage}[h]{1.0\linewidth}
  \begin{center} \epsfig{file=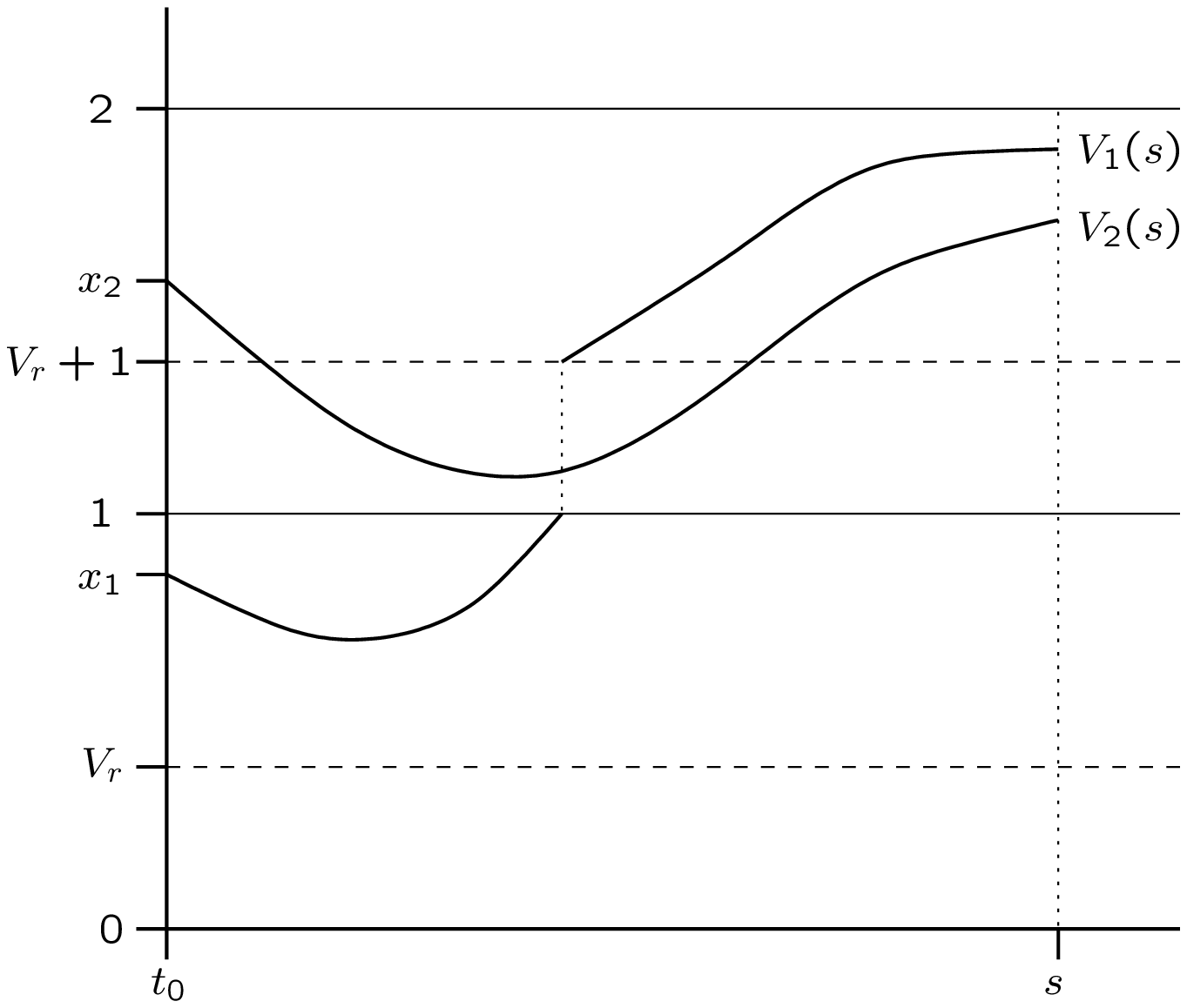, width=0.6\linewidth}
  \end{center} {\caption{\small Lack of monotonicity of the lift of the
      potential flow. Two solutions $V_1(t)$ and $V_2(t)$ with $V_1(t_0)=x_1 <
      x_2=V_2(t_0)$ have changed order at time $s$. \label{fig3}}}
\end{minipage}
\end{figure}

The most subtle issue in the proof of Theorem~\ref{t.asymptotic-ff} is the fact
that when $F$ takes negative values, then this might lead to discontinuities in
the flow $V(t)$ and to a lack of monotonicity with respect to the initial
condition $x_0$ (see Figure~\ref{fig3}). At this point, the monotonicity of $F$
in $V$ is needed in order to recover a certain {\em `almost-monotonicity'}
property. The discontinuity does not present a problem in the setting of RDS,
but impedes drawing further conclusions in the situation where the base flow
$\omega$ is a uniquely ergodic%
\foot{A map or flow on a compact metric space is called uniquely ergodic, if it
  has exactly one invariant probability measure.}
system on a compact metric space $\Theta$. All these complications can be
avoided by assuming that $F$ is non-negative. In this case, we have the
following.
\begin{thm}
  \label{t.aff-nonnegative} Suppose that the metric DS $(\Theta,{\cal
    B},\mu,\omega)$ is ergodic and the function $F$ in (\ref{e.Ftilde}) is
  bounded on $\Theta\times[0,1]$, uniformly Lipschitz-continuous in $V$ and
  $F(\theta,V_r) > 0 \ \forall \theta\in\Theta$. Then the following holds.
  \alphlist
\item The model described by (\ref{e.setback}) and (\ref{e.Ftilde}) has a unique
  asymptotic firing frequency: There exists a real number $\nu$ such that for
  $\mu$-almost every $\theta_0\in\Theta$ and all $x_0\in [V_l,V_u)$ there holds
  \begin{equation} \label{e.nu-convergence2}
    \nLim t_n/n \ = \ \nu \ .
  \end{equation}
\item If in addition $\Theta$ is a compact metric space, $F$ is continuous,
  $F(\theta,V_u)>0\ \forall \theta\in\Theta$ and $\omega$ is uniquely ergodic,
  then (\ref{e.nu-convergence2}) holds for all initial conditions
  $(\theta_0,x_0) \in \Theta\times [V_l,V_u)$ and the convergence is uniform in
  $(\theta_0,x_0)$.  \listend
\end{thm}
This statement can, for instance, be applied to the so-called {\em quadratic
  IFM}, which is given by (\ref{e.setback}) and (\ref{e.Ftilde}) with
\begin{equation}
  \label{e.quadratic-IFM}
  F(\theta,V) \ = \ V^2 + I(\theta) \ ,
\end{equation}
under the additional assumption that $\inf_{\theta\in\Theta} I(\theta) > \max\{-
V_r^2,-V_u^2\}$. (See \cite{brette:2004} for the analysis of this model in the
periodically forced case and further references).  The standard example of a
uniquely ergodic base flow would be the Kronecker flow $\theta\cdot t =
\theta+t(\omega_1\ld \omega_d)$ with $d$ rationally independent frequencies
$\omega_1\ld \omega_d$, corresponding to the excitation of ${\cal N}_1$ by $d$
independent pacemaker neurons. Although we will not discuss the topic in detail,
we want to mention that mode-locking phenomena may appear in this setting
whenever the asymptotic firing frequency is rationally related to the driving
frequencies $(\omega_1\ld \omega_d)$ \medskip

{\bf Acknowledgments.} The author would like to thank K.~Pakdaman for
interesting discussions that initiated this research. This work was supported by
a research fellowship (Ja 1721/1-1) of the German Research Council (DFG).

\section{Periodic forcing revisited} \label{Revisited}

In this section we briefly recall the analysis of the periodically forced IFM
and in particular the construction of the circle map $g$ mentioned in the
introduction. This will also allow to discuss the modifications needed to extend
the approach to randomly forced models. In order to simplify notation, we will
from now on always assume that $V_l=0$ and $V_u=1$.

Suppose that the function $F$ in (\ref{e.potential-law}) is $p$-periodic in the
second variable, that is
\begin{equation}
\label{e.F-periodicity}
F(t,V+p)\ = \ F(t,V) \ .
\end{equation}
Equivalently, we may assume that $\Theta = \kreis$ and $\omega_t(\theta) =
\theta+t/p \bmod 1$ in (\ref{e.Ftilde}). For simplicity, we also suppose that
\begin{equation} \label{e.simple-model}
  F(t,V_r) \ > \  0 \quad \forall t\in\R \ .
\end{equation}
In this case, we may assume without loss of generality that $V_r = V_l = 0$,
since the interval $(V_l,V_r)$ is not accessible from outside and does not play
a role in the description of the dynamics. The ODE (\ref{e.potential-law})
generates a flow
\begin{equation}
\Phi \ : \quad \R^2 \to \R \ , \quad (t,V) \mapsto \Phi_t(V)
\end{equation}
in the sense that $x(t) = \Phi_{t-t_0}(x_0)$ is the solution of
(\ref{e.potential-law}) with initial conditions $x(t_0)=x_0$. The first firing
time $t_1$ can then be expressed as $t_1 = t_0 + \inf\{t\geq 0 \mid
\Phi_{t-t_0}(x_0) = 1\}$.  Fixing the initial value $x_0=V_r$, this allows to
define a map $\tilde G : \R \to \R , \ t_0 \mapsto t_1$ that generates the
sequence of spiking times $t_n$, that is, $t_{n+1}=\tilde G(t_n) \ \forall
n\in\N_0$. Rescaling $\tilde G$, we let $G : \kreis \to \kreis , \ x \mapsto
G(pt)/p$.  The periodicity assumption (\ref{e.F-periodicity}) implies that
$\tilde G(t+p) = \tilde G(t)+p$ and therefore $G(t+1) = G(t)+1$.  Consequently $
G$ is the lift of a circle map $g : \kreis \to \kreis$. Furthermore, using
(\ref{e.simple-model}) it is possible to show the map $G$ is continuous and
strictly monotonically increasing, such that $g$ is an orientation-preserving
circle homeomorphism. Such maps have a well-defined rotation number defined by
the limit
\begin{equation}
  \label{e.g-rotnum}
  \rho(G) \ = \ \nLim (G^n(t)-t)/n \ = \ \nLim (\tilde G^n(pt)/p-pt)/pn
\end{equation}
which always exists and is independent of $t$ (see, for example,
\cite{katok/hasselblatt:1997}). Of course, this immediately implies the
existence and uniqueness of the asymptotic firing frequency $\nu_{{\cal N}_1} =
\frac{1}{p} \cdot \rho(\tilde G)^{-1}$.  \smallskip

There is also an alternative way of constructing the map $g$ that yields some
additional insight and which we want to discuss on an informal level. Suppose we
let
\begin{equation}
  \label{e.torus-vectorfield}
  \tilde F \ : \R^2 \to \R^2 \ , \quad (y,V) \mapsto (1/p,F(py,V-n))
   \quad \textrm{ if } V \in [n,n+1) \ .
\end{equation}
Then $\tilde F$ defines a vector field on $\R^2$ that is invariant under integer
translations and hence projects to a vector field on the two-torus
$\T^2$. Consequently, the flow $\tilde \Psi : \R \times \R^2 \to \R^2$ generated
by $\dot Y(t) = F(Y(t)),\ Y\in\R^2$, projects to a flow $\Psi$ on $\T^2$.  It is
now easy to see that the map $g$ defined above is simply the return map of
$\Psi$ to the Poincar\'e section $\T^1 \times \{0\}$. (Note that
(\ref{e.simple-model}) ensures that this section is transversal to the flow
direction.) However, from this point of view we also see that taking this
particular Poincar\'e section has something arbitrary. For example, one might
just as well have chosen to define another circle homeomorphism $\hat g$ with
lift $\hat G : \R \to \R$ by taking the Poincar\'e map of the vertical section
$\{0\} \times \T^1$. This simply corresponds to stopping the flow at times $\hat
t_n = np$. The asymptotic firing frequency could thus be obtained as $\nu_{{\cal
    N}_1} = \rho(\hat G) / p$. One can even arrive at the same conclusion
without resort to Poincar\'e sections at all.  As $\tilde F$ is constant in the
first coordinate, the flow $\Psi$ has skew product structure over the simple
base flow $\omega_t(t_0) = t_0 + t/p \ \bmod 1$. In other words, $\Psi$ is a
periodically forced circle flow with lift $\tilde \Psi$. As circle
homeomorphisms, such flows have a well-defined vertical rotation number and one
obtains $\nu_{{\cal N}_1} = \rho(\tilde \Psi)/p = \lim_{t\to\infty} \pi_2 \circ
\tilde\Psi_t(t_0,x_0)$.

Now, in the case of periodic forcing all this is surely a mere
tautology. However, things become quite different as soon as one wants to
consider more general types of forcing. When the driving space $\Theta$ is an
arbitrary measurable space, then taking a vertical Poincar\'e section does not
make sense anymore, whereas the Poincar\'e return map to $\Theta\times\{0\}$
just yields a self-map of $\Theta$ that might be very difficult to analyse. In
this context, the fact that we can also directly consider the forced circle flow
$\Psi$ generated by (\ref{e.setback}) and (\ref{e.Ftilde}) turns out to be a
great advantage. In the next section, we will present a result on randomly
forced circle flows and their lifts that ensures the existence and uniqueness of
the {\em vertical} or {\em fibred} rotation number of such skew product
flows. The remaining sections are then dedicated to the construction of suitable
lifts for the potential flow in the situation of Theorems \ref{t.asymptotic-ff}
and \ref{t.aff-nonnegative}, which immediately leads to proofs of the respective
statements.  \smallskip

Finally, before turning to the rigorous analysis in the next sections, we
briefly want to indicate why further technical problems appear when
condition~(\ref{e.simple-model}) does not hold anymore.  First of all, this
evidently leads to discontinuities of the flow when the potential `jumps' from
$V_u=1$ to $V_r \in (0,1)$ or, when considering lifts, from $V_u+n$ to $V_r+n+1$
for some $n\in\Z$. This would in itself not present a serious problem, since
most results on the existence and uniqueness of rotation numbers only require
monotonicity, whereas continuity is less important. However, it turns out that
dropping condition (\ref{e.simple-model}) may at the same time lead to a lack of
monotonicity of the circle flow. An example of how this can happen is indicated
in Figure . This remains true even when $F$ is non-increasing in $V$, as assumed
in Theorem~\ref{t.asymptotic-ff}. However, in this situation it is nevertheless
possible to make the argument work. In the case of periodic forcing, one can
show that lift $G$ of the circle map $g$ constructed above is monotonically
increasing on the image of one of its iterates (as done in
\cite{pakdaman:2001,brette:2004}). On the level of forced flows, one may
similarly show that while two different orbits may reverse their order, they
will always remain within a bounded distance of each other. (This is the point
of view we will adopt in Section~\ref{LIFM}.) In both cases, this is still
sufficient to guarantee the existence and uniqueness of the rotation number.

\section{Uniqueness of the fibred rotation number} \label{RDS-Rotnum}

As mentioned before, we say $(\Theta,{\cal B},\mu,\omega)$ is a metric DS if
$(\Theta,{\cal B},\mu)$ is a probability space and $\omega : \R \times \Theta
\to \Theta,\ (t,\theta) \mapsto \omega_t(\theta)$ is a flow that preserves
$\mu$, meaning that $\mu\circ \omega_t^{-1}=\mu \ \forall t\in\R$. As before, we
write $\omega_t(\theta) = \theta\cdot t$. A {\em random dynamical system (RDS)}
over $\omega$ is a measurable mapping
\begin{equation}\label{e.rds}
  \Psi \ : \quad \R \times
  \Theta\times\R \to \Theta\times\R \ , \quad (t,\theta,x) \mapsto (\theta\cdot t,\psi_t(\theta,x))
\end{equation}
that satisfies the {\em cocycle property}
\begin{equation}
  \label{e.cp}
  \psi_{t_1+t_2}(\theta,x) \ = \ \psi_{t_2}(\theta\cdot t_1,\psi_{t_1}(\theta,x))
  \quad \forall (t_1,t_2,\theta,x) \in \R^2 \times \Theta\times\R \  . 
\end{equation}
We say $\Psi$ is a {\em continuous RDS} if for all $\theta\in \Theta$ the
mapping $(t,x) \mapsto \psi_t(\theta,x)$ is continuous. 

Evidently, what we are interested in is the existence of uniqueness of the {\em
  fibred rotation number}
\begin{equation}
  \label{e.fibred-rotnum}
  \rho(\Psi) \ = \ \lim_{t\to\infty} (\psi_t(\theta,x)-x)/t \ 
\end{equation}
when $\Psi$ is obtained as the lift of a randomly forced circle flow. For this,
we will use the following
assumptions: \begin{list}{$\bullet$}{\usecounter{enumi}}
\item There holds
  \begin{equation} \label{e.A-circle-lift} \psi_t(\theta,x+1) \ = \
    \psi_t(\theta,x)+1 \quad \forall (t,\theta,x) \in \T \times \Theta \times \R \ .
\end{equation}
In particular, this implies that $\Psi$ projects to a random dynamical system on
the circle $\kreis$.
\item There exists a constant $C\geq 0$ such that 
  \begin{equation}\label{e.A-almost-monotonicity}
    \psi_t(\theta,x) \ \leq \ \psi_t(\theta,x') + C \quad \forall (t,\theta,x,x')
    \in \T \kreis \Theta\times \R^2 \textrm{ with } x \leq x' \ . 
\end{equation}
In other words, the mappings $\psi_t(\theta,\ldot)$ are {\em `almost
  monotone'} in $x$, up to a uniform constant $C$.
\item There exists a constant $\eta\geq 0$ such that
  \begin{equation} \label{e.A-uniform-continuity} |\psi_t(\theta,x)-x| \
    \leq \ \eta \quad \forall (t,\theta,x) \in [0,1]\times\Theta\times\R \ .
 \end{equation}
\listend

\begin{thm} \label{t.fibred-rotnum}
  Suppose that $(\Theta,{\cal B},\mu,\omega)$ is ergodic and the RDS $\Psi$
  satisfies (\ref{e.A-circle-lift})--(\ref{e.A-uniform-continuity}).  Then there
  exists a real number $\rho$ and a set $\Theta_0 \ssq \Theta$ of full measure,
  such that
\begin{equation}
    \label{eq:3}
    \lim_{t\to\infty} (\psi(t,\theta,x)-x)/t \ = \ \rho \quad 
    \forall (\theta,x) \in \Theta_0\times \R \  .
\end{equation}
\end{thm}
\begin{rem}
  \alphlist
\item When $\Theta$ is a compact metric space and $\omega$ a uniquely ergodic
  flow on $\Theta$, then this result is well-known and due to Herman
  \cite{herman:1983}. (See also \cite{johnson/moser:1982} for a precursor in the
  context of quasiperiodic Schr\"odinger operators.) For the more general case
  of ergodically forced monotone circle maps, it was proved more recently by Li
  and Lu \cite{li/lu:2006}. The only new aspect here is that we work with
  continuous time replace monotonicity by the slightly weaker property
  (\ref{e.A-almost-monotonicity}). It is not surprising that this can be proved
  along the same lines as the previous results, but since the proof is rather
  short anyway we include the details for the convenience of the reader.
\item When $\Psi$ is a continuous RDS on the circle, then assumption
  (\ref{e.A-almost-monotonicity}) can be replaced by the much more general one
that
\begin{equation}
  \lim_{t\to\infty} |\psi_t(\theta,x) - \psi_t(\theta,x')|/t \ = 
  \ 0 \quad \forall (\theta,x,x') \in \Theta\times \R^2 \ . 
\end{equation}
In this case Herman's original proof, which uses the existence of a
$\Psi$-invariant probability measure that projects down to $\mu$, remains valid
with hardly any modifications. However, since we do not assume $\Psi$ to be
continuous (and this is crucial for the application to the forced LIFM), such an
invariant measure does not necessarily exist. \listend
\end{rem}

{\em Proof of Theorem~\ref{t.fibred-rotnum}.} By (\ref{e.A-circle-lift}) and
(\ref{e.A-almost-monotonicity}) we have
\begin{equation}
  \label{eq:4} |(\psi_t(\theta,x)-x) - (\psi_t(\theta,x')-x')| 
  \ \leq \ C+1 \quad \forall (t,\theta)\in \T\times\Theta, x,x'\in A_\theta \ .
\end{equation}
(Note that due to (\ref{e.A-circle-lift}) we may assume w.l.o.g.\ that $x \in
[x'-1,x')$), in which case the estimate is a direct consequence of
(\ref{e.A-almost-monotonicity}).) It follows that when the limit $\rho(\theta,x)
:= \lim_{t\to\infty}(\psi_t(\theta,x)-x)/t$ exists for one $x\in\R$, it
exists for all $x\in\R$ and does not depend on $x$.  Given any $n\in\N_0$, let
$F_n(\theta) := \psi_n(\theta,0) + C + 1$.  Then (\ref{eq:4}) implies that
\begin{eqnarray*}
  F_{n+m}(\theta) & =  & \psi_{n+m}(\theta,0) +C +1 \
  = \  \psi_m(\theta\cdot n,\psi_n(\theta,0)) -
  \psi_n(\theta,0) + F_n(\theta)  \\
  & \stackrel{(\ref{eq:4})}{\leq} & \psi_m(\theta\cdot n,0) + C + 1 + F_n(\theta) \ = \
  F_m(\theta\cdot n) + F_n(\theta) \ .
\end{eqnarray*}
Thus the random variables $F_n$ form a subadditive sequence over the
measure-preserving transformation $\theta \mapsto \theta\cdot 1$. Further
(\ref{e.A-uniform-continuity}) implies that $F_1$ is bounded. Hence, we can
apply Kingman's Subadditive Ergodic Theorem (see, for example,
\cite{arnold:1998}), which yields the existence of a $\mu$-integrable function
$\tilde\rho : \Theta \to \R$ and a set $\Theta' \ssq \Theta$, such that for all
$\theta \in \Theta'$ there holds
\[
\nLim F_n(\theta)/n \ = \ \tilde\rho(\theta)  \ .
\]
(\ref{e.A-uniform-continuity}) and (\ref{eq:4}) together now now imply that for
all $(\theta,x)\in\Theta'\times\R$ we have 
\[
\lim_{t\to\infty} (\psi_t(\theta,x)-x)/t \ = \ \tilde\rho(\theta) \ .
\]
It is easy to see that the function $\tilde\rho$ is invariant, that is
$\tilde\rho(\theta\cdot t) = \tilde\rho(\theta) \ \forall t\in\T$, and the
ergodicity of the base flow $\omega$ therefore implies that $\tilde\rho(\theta)
= \int_\Theta \tilde\rho \ d\mu =: \rho$ on a set of full measure $\Theta''$. If
we now let $\Theta_0 = \Theta' \cap \Theta''$ then all the assertions of the
theorem are satisfied.  \qed

\section{Construction of the potential flow} \label{Construction}

The aim of this section is to formalise the model described by (\ref{e.setback})
and (\ref{e.Ftilde}) in the introduction and to prove
Theorem~\ref{t.aff-nonnegative}. (The proof of Theorem~\ref{t.asymptotic-ff}
will then be given in Section~\ref{LIFM}.) More precisely, we will construct a
lift $\wh V$ for the circle flow $V$ that describes the evolution of the
membrane potential (recall that we identify the interval $[V_l,V_u)$ with the
circle. We assume without loss of generality that $V_l=0$ and \mbox{$V_u=1$.}
Unfortunately, the construction has a somewhat technical flavour, which
basically comes from the need to treat the discontinuities produced by
(\ref{e.setback}) in a formally precise way. However, in order to treat the
problem in a rigorous way a certain amount of detail seems unavoidable, in
particular since the {\em `almost-monotonicity property'} needed for the proof
of Theorem~\ref{t.asymptotic-ff} is a rather subtle issue.  \medskip

Suppose that $(\Theta,{\cal B},\mu,\omega)$ is a metric DS and consider the
non-autonomous differential equation
\begin{equation}
  \label{e.generator}
  \dot x(t) \ = \ F(\theta\cdot t,x(t))
\end{equation}
with $F:\R \times \Theta \to \R$, which is equivalent to (\ref{e.Ftilde}). As
mentioned in the introduction, we first discuss the precise technical conditions
on $F$ needed for the construction. Given any $f:\R \to \R$, we let
\begin{eqnarray*}
  \| f \|_{[0,1],0,0} & = & \sup_{x\in[0,1]} |f(x)| \quad \textrm{ and } \\
  \| f \|_{[0,1],0,1} & = & \| f\|_{[0,1],0,0} + 
  \sup_{x\neq y\in[0,1]} \frac{|f(y)-f(x)|}{|y-x|} \ .
 \end{eqnarray*}
 (The terminology follows \cite[Appendix B]{arnold:1998}.) We then require that
 there exists a constant $C>0$, such that for all $\theta\in\Theta$ there holds
\begin{eqnarray}
  \int_0^1 \| F(\theta\cdot t,\ldot)\|_{[0,1],0,0} \ dt & \leq & C \quad \textrm{ and } 
  \label{e.bounded-firing} \\
  \int_0^1 \| F(\theta\cdot t,\ldot)\|_{[0,1],0,1} \ dt & < & \infty \ .
  \label{e.generator-condition}
\end{eqnarray}
The restriction to the interval $[0,1]$ in the definition of the norms is
explained by the fact that due to the reset procedure described by
(\ref{e.setback}) this is the only part of the phase space we are interested
in. We could equally assume that the conditions are satisfied on all of $\R$ and
just modify the function $F$ if they are violated outside of $[0,1]$. Assumption
(\ref{e.bounded-firing}) is needed to ensure that the hypothesis of
Theorem~\ref{t.fibred-rotnum} in the appendix are met -- roughly spoken, it just
ensures that action potentials or {\em spikes} cannot be generated at an
arbitrary rate (there will be at most $C/(1-V_r)+1$ spikes in any interval of
length 1). (\ref{e.generator-condition}) is just a standard technical condition
which ensures that (\ref{e.generator}) generates a RDS
\begin{equation}
  \Phi \ :\quad \R \times \Theta \times \R \ \to \Theta \times \R \ , 
  \quad \Phi(t,\theta,x) = (\theta\cdot t,\varphi_t(\theta,x))
\end{equation}
over the base flow $\omega$ (see \cite[Theorem~2.2.2]{arnold:1998}).%
\foot{Strictly spoken, we would have to take into account finite escape times,
  such that $t\mapsto \varphi_t(\theta,x)$ is only defined on a subinterval of
  $\R$. However, since we are only interested in the dynamics on
  $\Theta\times[0,1]$ we assume for simplicity that (\ref{e.generator}) always
  has global solutions.}  From now on, we will always assume that $\Phi$ is the
skew product flow generated by (\ref{e.generator}).\medskip

In the following, we will give a precise definition of the {\em lift of the
  potential semiflow}
\begin{equation}
  \label{e.voltage-flow}
  \wh V \ : \quad \R^+ \times \Theta \times \R \ , \quad (t,\theta,v) 
  \mapsto (\theta\cdot t,\wh V_t(\theta,x)) \ 
\end{equation}
corresponding to the model described by (\ref{e.setback}) and
(\ref{e.Ftilde}). The membrane potential at time $t$ with initial values $t=0$,
$\theta$ and $x_0$ is then obtained as $V(t) = \wh V_t(\theta,x_0) \bmod 1$. The
only assumption on $F$ needed for this construction is
(\ref{e.generator-condition}). (\ref{e.bounded-firing}) and the additional
assumptions made in Theorems~\ref{t.asymptotic-ff} and \ref{t.aff-nonnegative}
will be required only later in order to ensure that $\widehat V$ meets the
assumptions of Theorem~\ref{t.fibred-rotnum}.

Given any $x\in\R$, we denote by $[x]\in\Z$ its integer part and by
$\{x\}=x-[x]\in[0,1)$ its fractional part. For any $(\theta,x) \in
\Theta\times\R$ we define
\begin{equation}
  \label{e.tau}
  \tau(\theta,x) \ = \ \inf\left\{t\geq 0\mid \varphi_t(\theta,\{x\}) 
  = 1 \right\}  \ 
\end{equation}
with $\tau(\theta,x) = \infty$ if the set on the right is empty. This is the
time when the first spike is generated.

For any $s\in (\R^+)^\N$ we denote by $S=S(s)$ the sequence given by
$S_n=\sum_{i=0}^n s_i$. Then, given $(\theta,x) \in \Theta\times\R$, we
recursively define a sequence $s_n=s_n(\theta,x)$ and the corresponding sequence
$S_n=S_n(\theta)$ by
\begin{equation}
  \label{e.sn}
  s_0 = 0 \ , \ s_1 = \tau(\theta,x) \quad \textrm{and} \quad s_{n+1} = \tau(\theta\cdot S_n,V_r) \ . 
\end{equation}
(Recall that $V_r \in [0,1)=[V_l,V_u)$ is the rest potential introduced in
(\ref{e.setback}).) $S_n$ is the time when the $n$-th action potential is
triggered, whereas $s_n$ is the length of the time interval between the $n-1$-th
and the $n$-th spike.
Given $(t,\theta,x)\in \R^+\times\Theta\times \R$ we define
$n(t,\theta,x)$ as the unique integer $n$ such that $S_n(\theta,x) \leq t <
S_{n+1}(\theta,x)$, that is 
\begin{equation}
     n(t,\theta,x) \ = \ \max\{n\in\N \mid S_n(\theta,x) \leq t \} \ .
\end{equation}
In other words, $n(t,\theta,x)$ is just the number of spikes generated until
time $t$. The lifted potential flow is then defined by (\ref{e.voltage-flow})
with
\begin{equation}
  \label{e.vf-definition}
  \wh V_t(\theta,x) \ = \ \left\{
    \begin{array}{ll}
      [x] + \varphi_t(\theta,\{x\}) & \textrm{if } n(t,\theta,x) = 0 \ ;   \\ \ \\
      {[}x] + n + \varphi_{t-S_n}(\theta\cdot S_n,V_r) &  \textrm{if } n(t,\theta,x) = n \ . 
    \end{array} \right.
\end{equation}
\begin{rem}\label{r.vf-definition}\alphlist
 \item
  The circle flow $V$ defined by (\ref{e.setback}) and (\ref{e.Ftilde}) is
  typically not continuous and therefore {\em a priori} has many different lifts
  that do not only differ by an integer constant. However, the particular lift
  $\wh V$ defined above is the unique one that `counts' the number
  $n(t,\theta,x)$ of spikes (action potentials) generated up to time $t$, in the
  sense that $n(t,\theta,x)$ equals the number of integers in the interval
  $[x,\wh V_t(\theta,x)]$. In this way, the asymptotic firing frequency is
  obtained as the inverse of the rotation number
  \begin{equation} \label{e.Vt}
    \rho(\wh V) \ = \ \lim_{t\to \infty} (\wh V_t(\theta,x))/t \ ,
  \end{equation}
  provided this limit exists. To show that this is the case under the
  assumptions made in Theorem~\ref{t.aff-nonnegative} is the aim of the
  remainder of this section.
\item We note that 
  \begin{equation}
    \label{e.vf-periodicity}
    \wh V_t(\theta,x+1) \ = \ \wh V_t(\theta,x) +1 \quad \forall (t,\theta,x)\in 
  \R^+\times \Theta\times \R \ .
  \end{equation}
Further, when (\ref{e.bounded-firing}) holds then 
\begin{equation}
  \label{e.vf-bounded-firing}
  |\wh V_t(\theta,x) - x | \ \leq \ \frac{C}{1-V_r} +1 \quad 
  \forall t\in[0,1],\ (\theta,x) \in \Theta\times \R \ . 
\end{equation}
Finally, we note that by definition we have 
\begin{equation}
  \label{e.vf-right-continuity}
\wh  V_{S_n}(\theta,x) \ = \ [x] + n + V_r \ .
\end{equation}
Consequently the mapping $t\mapsto \wh V_t(\theta,x)$ is continuous from the
right. The only discontinuities occur exactly at times $t=S_n$, except in the
case $V_r=0$ in which $t \mapsto \wh V_t(\theta,x)$ is continuous. \listend
\end{rem}
   
\begin{prop}
  \label{p.vf-properties}
  The mapping $\wh V$ given by (\ref{e.voltage-flow}) and (\ref{e.Vt})
  defines a RDS over the base flow $\omega$. In particular, it has the cocycle
  property
  \begin{equation}
    \label{e.cocycle}
    \wh V_{t_1+t_2}(\theta,x) \ = \ \wh V_{t_2}(\theta\cdot t_1,V_{t_1}(\theta,x)) 
    \quad \forall t_1,t_2\in\R^+,\ (\theta,x) \in \Theta\times\R\  . 
  \end{equation}
  Furthermore, if $V_r = 0$ and $F(\theta,0)>0 \ \forall \theta\in\Theta$ then
  the mapping $x \mapsto V_t(\theta,x)$ is monotonically increasing for all
  $(t,\theta)\in\R^+$. If in addition $\omega$ is a continuous flow on a
  topological space $\Theta$, $F$ in (\ref{e.generator}) is continuous and
  $F(\theta,1)>0\ \forall \theta\in\Theta$, then $\wh V$ is continuous as a
  mapping $\R^+\times\Theta\times\R \to \Theta \times \R$.
\end{prop}
This shows that in the situation of Theorem~\ref{t.aff-nonnegative} the
semi-flow $\wh V$ satisfies all assumptions of Theorem~\ref{t.fibred-rotnum},
which proves part (a) of the theorem. Part (b) for uniquely ergodic base flows
then follows from Herman's classical result on the fibred rotation number
\cite[Theorem 5.4]{herman:1983}. \medskip

\myproof According to the definition of a RDS, we have to show that the mapping
$\wh V$ is measurable and has the cocycle property. (Since measurability follows
from standard arguments we omit the details.) \medskip

{\em Cocycle property.} In order to show (\ref{e.cocycle}) we need the following technical
statement.
\begin{claim}
  \label{c.vf-properties} Given $t_1,t_2\geq 0$ and $(\theta,x) \in \Theta
  \times \R$, let $n_1=n(t_1,\theta,x)$, $\sigma_1=S_{n_1}(\theta,x)$,
  $n_2=n(t_2,\theta\cdot t_1,\varphi_{t_1-\sigma_1}(\theta\cdot\sigma_1,V_r))$,
  $\sigma_2=S_{n_2}(\theta\cdot t_1,\varphi_{t_1-\sigma_1}(\theta\cdot
  \sigma_1,V_r))$ and $n = n(t_1+t_2,\theta,x)$. Then
  \begin{eqnarray}
    \label{e.claim1} n & = & n_1+n_2 \quad \textrm{ and} \\
    \label{e.claim2} S_n(\theta,x)&=& t_1+\sigma_2 \ . 
  \end{eqnarray}
\end{claim}

\myproof First, we note that $n_1=n(t_1,\theta,x) = n(\sigma_1,\theta,x)$ and
$n_2=n(t_2,\theta\cdot \sigma_1,V_r)$. Hence, we may assume without loss of
generality that $t_1=\sigma_1$. We claim that 
\begin{equation} \label{e.claim-induction}
S_k(\theta\cdot S_j(\theta,x),V_r)\  = \ S_{k+j}(\theta,x) - S_j(\theta,x) \quad
\forall j,k\in\N \ .
\end{equation}
In order to see this, we proceed by induction on $k$. There holds
\[
S_1(\theta\cdot S_j(\theta,x),V_r) \ = \ \tau(\theta\cdot S_j(\theta,x),V_r) =
s_{j+1}(\theta,x) \ = \ S_{j+1}(\theta,x)-S_j(\theta,x) \ .
\]
Further, if (\ref{e.claim-induction}) holds for $k$, then 
\begin{eqnarray*}
  S_{k+1}(\theta\cdot S_j(\theta,x),V_r) &= & 
  \tau(\theta\cdot (S_j(\theta,x)+S_k(\theta\cdot S_j(\theta,x),V_r)),V_r)
  + S_k(\theta\cdot S_j(\theta,x),V_r)  \\
  & = & \tau(\theta\cdot S_{k+j}(\theta,x),V_r) + S_{k+j}(\theta,x) - S_j(\theta,x) \\
  & = & s_{k+j+1}(\theta,x) + S_{k+j}(\theta,x) - S_j(\theta,x) \ 
  = \ S_{k++j+1}(\theta,x) - S_j(\theta,x) \ . 
\end{eqnarray*}
This proves (\ref{e.claim-induction}). Now, we have
\begin{eqnarray*}
  \lefteqn{n(t_2,\theta\cdot\sigma_1,V_r) \ = \
    \max\{k\in\N \mid S_k(\theta\cdot \sigma_1,V_r) \leq t_2\} }\\
  &\stackrel{(\ref{e.claim-induction})}{=} & 
  \max\{k\in\N \mid S_{n_1+k}(\theta,x)-\sigma_1 \leq t_2\}   \\ &
  = & \max\{k\in\N \mid S_{n_1+k}(\theta,x) \leq t_2 + \sigma_1\} \\
  & = & \max\{S_k(\theta,x) \leq t_2 + \sigma_1\} - n_1 \  = \ n(t_2+\sigma_1,\theta,x) - n_1 \ .
\end{eqnarray*}
Thus (\ref{e.claim1}) holds. In order to show (\ref{e.claim2}), recall that we
assumed $t_1=\sigma_1$. Therefore
\[
  t_1 + \sigma_2 
 \  = \  \sigma_1+S_{n_2}(\theta\cdot\sigma_1,V_r) 
  \ \stackrel{(\ref{e.claim-induction})}{=} \ S_{n}(\theta,x) \ .
\]
\roundqed \medskip

Now the cocycle property follows easily. Given $t_1,t_2\geq 0$, we define
$n_1,n_2,\sigma_1$ and $\sigma_2$ as in the claim above. First, suppose that
$n_1,n_2\geq 1$. Then
\begin{eqnarray*}
  \wh V_{t_2}(\theta\cdot t_1,\wh V_{t_1}(\theta,x)) 
  & \stackrel{(\ref{e.vf-periodicity})}{=} & 
  [x]+n_1+\wh V_{t_2}(\theta\cdot t_1,\varphi_{t_1-\sigma_1}(\theta\cdot \sigma_1,V_r)) \\
  & = & [x]+n_1+n_2+\varphi_{t_2-\sigma_2}(\theta\cdot(t_1+\sigma_2),V_r) \\
  & \stackrel{(\ref{e.claim2})}{=} & [x]+n+\varphi_{t_1+t_2-S_n(\theta,x)}(\theta\cdot S_n(\theta,x),V_r) 
  \ = \ \wh V_{t_1+t_2}(\theta,x) \ .
\end{eqnarray*}
The cases where $n_1=0$ or $n_2=0$ are treated similarly.

{\em Continuity and monotonicity.} Suppose that $F$ is strictly positive and
$V_r=0$. As mentioned in Remark~\ref{r.vf-definition}, the mapping $(t,x)
\mapsto \wh V_t(\theta,x)$ is continuous in $t$ in this case. In order to show
the monotonicity, fix $\theta\in\Theta$ and $x_1<x_2 \in \R$ and let $t_0 =
\inf\{t\in\R^+ \mid \wh V_t(\theta,x_1) \geq \wh V_t(\theta,x_2)\}$.

Suppose for a contradiction that $t_0 < \infty$. Then continuity in $t$ yields
$V_0 := \wh V_{t_0}(\theta,x_1) = \wh V_{t_0}(\theta,x_2)$. We distinguish two
cases. First, assume that $V_0 \notin \Z + V_r$. In this case the orbits of
$(\theta,x_1)$ and $(\theta,x_2)$ coincide with integer translates of orbits of
the flow $\Phi$ on a small interval $I_0$ around $t_0$. They are therefore
either equal or distinct on all of $I_0$, but cannot merge exactly at time
$t_0$.  Secondly, assume that $V_0 = k + V_r$ for some $k\in\Z$. Since $F$ is
strictly positive, this would imply that $\lim_{t\to t_0} \wh V_t(\theta,x_1) =
\lim_{t\to t_0} \wh V_t(\theta,x_2) = k$, which would again mean that two
distinct orbits of the flow $\Phi$ would have to merge at time $t_0$. Hence, in
both cases we arrive at a contradiction.

Now assume in addition that $\omega$ is a continuous flow on a topological space
$\Theta$, $F$ is continuous and 
\begin{equation} \label{e.F-positivity} F(\theta,1)\ > \ 0 \quad \forall
  \theta\in\Theta \ .
\end{equation}
Then $F$ is bounded on $\Theta\times[0,1]$, which together with the uniform
Lipschitz continuity of $F$ in $V$ implies that for any compact interval $I\ssq
\R$ the flow $\Phi$ generated by (\ref{e.generator}) is uniformly continuous on
$I\times\Theta\times[0,1]$. Now, the flow $\wh V$ is obtained by concatenating
integer translates of finite trajectories of $\Phi$ in $\Theta\times[0,1]$. $\wh
V$ will therefore inherit the uniform continuity of $\Phi$, provided that no
discontinuities are created by this concatenation in
(\ref{e.vf-definition}). 

In order to see this, observe that (\ref{e.F-positivity}) together with the
continuity of $F$ implies that $\tau$ defined in (\ref{e.tau}) is continuous in
$(\theta,x)$, except when $x\in\Z$. By induction, this yields that the spiking
times $S_n(\theta,x)$ depend continuously on $(\theta,x)$ as well unless
$x\in\Z$. Consequently $n(t,\theta,x)$ is continuous in $(t,\theta,x)$ unless
$x\in\Z$ or $t=S_n(\theta,x)$ for some $n\in\Z$. Furthermore $[x]$ is obviously
locally constant when $x\notin\Z$. Thus, it only remains to check that $\wh
V_t(\theta,x)$ defined by (\ref{e.vf-definition}) is continuous in
$(t,\theta,x)$ when $x\in\Z$ or when $t=S_n(\theta,x)$. However, this can be
seen quite easily by having a careful look at (\ref{e.vf-definition}). When $x$
passes an integer, then $[x]$ will jump up by one, but at the same time
$n(t,\theta,x)$ will drop down by one, such that the two discontinuities cancel
each other. Similarly, when $t$ and $S_n(\theta,x)$ change order then
$n(t,\theta,x)$ has a discontinuity of size $1$, but at the same time the
quantity $\varphi_{t-S_{n(t,\theta,x)}}(\theta\cdot
S_{n(t,\theta,x)}(\theta,x),V_r)$ jumps by one in the opposite direction. \qed
\medskip


\section{Modifications needed for the forced LIFM}\label{LIFM}

\begin{prop}
  \label{p.vf-almost-monotonicity}
  Suppose that the function $F$ in (\ref{e.generator}) is non-increasing in
  $x$. Then the mapping $\wh V$ defined by (\ref{e.voltage-flow}) and
  (\ref{e.vf-definition}) satisfies
  \begin{equation}
    \label{e.vf-almost-monotonicity}
    \wh V_t(\theta,x_1) \ \leq \ \wh V_t(\theta,x_2) + K 
    \quad \forall (t,\theta,x_1,x_2) \in \R^+\times\Theta\times\R^2 \textrm{ with } x_1\leq x_2 \ ,
  \end{equation}
  where $K = \frac{V_r}{1-V_r}+1$.
\end{prop}

\myproof We first prove that for all $(t,\theta) \in \R^+\times \Theta$ there
holds
\begin{equation}
  \label{e.vf-am1}
  \wh V_t(\theta,x_2) -1 \ \leq \ \wh V_t(\theta,x_1) \ 
  \leq \ \wh V_t(\theta,x_2)
  \quad \forall x_1,x_2 \in [V_r,1) \textrm{ with } x_1 < x_2 \ .
\end{equation}
Fix $\theta\in\Theta$ and $x_1< x_2 \in [V_r,1)$. Let $S_n := S_n(\theta,x_1)$
and $S_n' := S_n(\theta,x_2)$.  Then we show by induction on $n$ that
\begin{eqnarray}\label{e.vf-am2} 
  \wh V_t(\theta,x_2) -1 & \leq & \wh V_t(\theta,x_1) \ 
  \leq \ \wh V_t(\theta,x_2) 
  \quad \forall t \in [0,S_n'] \quad \textrm{ and}  \\ 
  \wh V_{S_n'}(\theta,x_1) & \in & (n-1+V_r,n) \ .  \label{e.vf-am3} 
\end{eqnarray}
In order to do so, we use the fact that for all $\theta\in\Theta$, $t>0$ and
$x\leq x'$ there holds
\begin{equation} \label{e.phi-contraction}
\varphi_t(\theta,x') - \varphi_t(\theta,x) \ \leq \  x'-x
\end{equation}
since
\[
\varphi_t(\theta,x') - \varphi_t(\theta,x) \ = \ x'-x + \int_{0}^t
F(\theta\cdot r,\varphi_r(\theta,x')) - F(\theta\cdot r,\varphi_r(\theta,x))\
dr
\]
and the integral on the right is non-positive since $F$ is non-increasing. In
order to start the induction, note that $\tau(\theta,x_1) \geq
\tau(\theta,x_2)$. Hence, if $t< S_1 =\tau(\theta,x_2)$, then
\begin{equation} \label{e.inductionstart} 0 \ < \ \wh V_t(\theta,x_1) \ =
  \varphi_t(\theta,x_1) \ < \ \varphi_t(\theta,x_2) \ = \ \wh
  V_t(\theta,x_2) \ < \ 1 \ .
\end{equation}
Further~(\ref{e.phi-contraction}), $x_2-x_1 < 1-V_r$ and the fact that
$\lim_{t\nearrow S_1'}\wh V_t(\theta,x_2) = 1$ imply that
\[
V_r \ < \ V_{S_1'}(\theta,x_1) \ < \ 1 \ .
\]
This shows that (\ref{e.vf-am2}) and (\ref{e.vf-am3}) hold for $n=1$.\smallskip

Now, suppose that (\ref{e.vf-am2}) and (\ref{e.vf-am3}) hold for $n\geq 1$. Then
using (\ref{e.phi-contraction}) and $\lim_{t\nearrow S_n}\wh
V_t(\theta,x_1) = n$ in order to compare $V_t(\theta,x_2)-1$ and
$V_t(\theta,x_1)$, similar as in (\ref{e.inductionstart}), we obtain that 
\begin{eqnarray*}
n-1 & < & V_t(\theta,x_2)-1 \ < \ V_t(\theta,x_1) \ < \ n \quad \forall t\in
[S_n',S_n) \quad  \textrm{ and} \\
n-1+V_r & < & V_{S_n}(\theta,x_2) -1 \ < \ n \ .
\end{eqnarray*}
This shows that (\ref{e.vf-am1}) holds on the interval $[S_n',S_n]$, and for the
remaining interval $(S_n,S_{n+1}']$ we can now proceed in exactly the same way
as in the case $n=0$. This proves (\ref{e.vf-am2}) and (\ref{e.vf-am3}) for all
$n\in\N$ and hence (\ref{e.vf-am1}). \smallskip

In order to treat the general case, now assume that $x_1<x_2$ are arbitrary. Due
to (\ref{e.vf-periodicity}) we may assume without loss of generality that $x_1
\in [0,1)$ and $x_2 \in (x_1,x_1+1)$. We distinguish two cases. First, assume
that $x_2 \in (x_1,1)$. Let $S_n' = S_n(\theta,x_2)$ as above and
\[
m \ := \ \max\{n\in\N \mid \wh V_{S_n'}(\theta,x_1) \in (0,V_r)\} \ .
\]
Then, since $\wh V_{S_m'}(\theta,x_1) = \varphi_{S_m'}(\theta,x_1)$ is still
below $V_r = \wh V_{S_m'}(\theta,x_2) - m$ we have that $\wh
V_{S_{m+1}'}(\theta,x_1)$ is still below $1$. Consequently both $\wh
V_{S_{m+1}'}(\theta,x_1)$ and $\wh V_{S_{m+1}'}(\theta,x_2)$ are contained in
$[V_r,1)$. Using the cocycle property (\ref{e.cocycle}) we can therefore apply
(\ref{e.vf-am1}) to see that 
\[
\wh V_t(\theta,x_1) - 1 \ \leq \ \wh V_t(\theta,x_2) -(m+1) \ \leq \ \wh
V_t(\theta,x_1) \quad \forall t \geq S_{m+1}' \ .
\]
We thus obtain 
\begin{equation}
  \label{e.am-firststep}
  \wh V_t(\theta,x_2) - (m+1) \ \leq \ \wh V_t(\theta,x_1) 
  \ \leq \wh V_t(\theta,x_2) \quad \forall t\in\R^+ \ . 
\end{equation}
This already proves (\ref{e.vf-almost-monotonicity}) for such
$x_1,x_2$. 

In order to treat the case where $x_2 \in (1,1+x_1)$ we have to obtain some
information on $m$. More precisely, we claim that
\begin{equation}
  \label{e.m}
  m \ \leq \ \frac{V_r}{1-V_r} + 1 \ .
\end{equation}
This follows from the fact that for all $n\leq m$ there holds
\[
\wh V_{S_n'}(\theta,x_1) \ \geq \ (n-1)\cdot (1-V_r)
\]
which can be proved easily by induction using (\ref{e.phi-contraction}) together
with the fact that $\wh V_{S_n'}(\theta,x_2)-n = V_r$ and $\lim_{t\nearrow
  S_{n+1}'}\wh V_t(\theta,x_2)-n = 1$.

Now assume that $x_2 \in (1,1+x_1)$. Then we can apply (\ref{e.am-firststep}) to
$x_1'=x_2-1$ and $x_2'=x_1$ to obtain that 
\[
\wh V_t(\theta,x_1) - (m+1) \ \leq \ \wh V_t(\theta,x_2) - 1 \quad \forall
t\in\R^+
\]
and hence
\[
\wh V_t(\theta,x_1) \ \leq \ V_t(\theta,x_2) + m \quad \forall t\in\R^+ \ .
\]
Together with (\ref{e.m}) this proves (\ref{e.vf-almost-monotonicity}). 
\qed




{


}
\end{document}

%% file: bibstandard.tex
\newcommand{\ldot}{\ensuremath{\textbf{.}}}
\newcommand{\ld}{\ensuremath{,\ldots,}}
\newcommand{\ssq}{\ensuremath{\subseteq}}

\newcommand{\T}{\ensuremath{\mathbb{T}}}

\newcommand{\N}{\ensuremath{\mathbb{N}}} 
\newcommand{\R}{\ensuremath{\mathbb{R}}}
\newcommand{\Z}{\ensuremath{\mathbb{Z}}}

\newcommand{\kreis}{\ensuremath{\mathbb{T}^{1}}}

\newcommand{\alphlist}{\begin{list}{(\alph{enumi})}{\usecounter{enumi}\setlength{\parsep}{2pt}
      \setlength{\itemsep}{1pt} \setlength{\topsep}{5pt}
      \setlength{\partopsep}{3pt}}}

\newcommand{\arablist}{\begin{list}{(\arabic{enumi})}{\usecounter{enumi}\setlength{\parsep}{2pt}
          \setlength{\itemsep}{1pt} \setlength{\topsep}{5pt}
          \setlength{\partopsep}{3pt}}}

\newcommand{\romanlist}{\begin{list}{(\roman{enumi})}{\usecounter{enumi}\setlength{\parsep}{2pt}
              \setlength{\itemsep}{1pt} \setlength{\topsep}{5pt}
              \setlength{\partopsep}{3pt}}}

 \newcommand{\listend}{\end{list}}

\newcommand{\bulletlist}{\begin{list}{$\bullet$}{\setlength{\parsep}{2pt}
                \setlength{\itemsep}{1pt} \setlength{\topsep}{5pt}
                \setlength{\partopsep}{3pt}\setlength{\leftmargin}{15pt}}}

\newcommand{\roundqed}{{\hfill \Large $\circ$}}

\newcommand{\myproof}{\textit{Proof. }}

\newcommand{\foot}{\footnote}

\newcommand{\nLim}{\ensuremath{\lim_{n\rightarrow\infty}}}

